\documentclass[aps,prx,twocolumn]{revtex4}
\usepackage[final]{pdfpages}
\usepackage{pgffor}

\usepackage{dcolumn}
\usepackage{graphicx}
\usepackage{bm}
\usepackage{color}
\usepackage{xcolor}
\usepackage{amsmath}
\usepackage{amstext}
\usepackage{amssymb}
\usepackage{booktabs}
\usepackage{array}
\usepackage[colorlinks=true,linkcolor=blue,citecolor=blue,urlcolor=blue]{hyperref}
\usepackage{ulem}
\usepackage{verbatim} 

\newcommand{\suppinfo}{Supplemental Material~\cite{supp-info}}

\renewcommand{\emph}[1]{\textit{#1}}

\newcommand{\editor}[2]{%
  \expandafter\newcommand\csname #1note\endcsname[1]{%
    \textcolor{#2}{(\textbf{#1:} {\it ##1})}}%
  \expandafter\newcommand\csname #1\endcsname[1]{%
    \textcolor{#2}{##1}}%
  \expandafter\newcommand\csname #1cancel\endcsname[1]{%
    \textcolor{#2}{\sout{##1}}}%
  \expandafter\newcommand\csname #1change\endcsname[2]{%
    \textcolor{#2}{\sout{##1} ##2}}%
  \newenvironment{#1text}{\color{#2}}{\color{black}}
}
\definecolor{Blu}{rgb}{0.00,0.00,1.00}
\definecolor{Red}{rgb}{1.00,0.00,0.00}
\definecolor{Cyan}{rgb}{0.00,0.50,0.50}
\definecolor{Green}{rgb}{0.00,0.70,0.00}
\definecolor{BluBondi}{rgb}{0.00,0.58,0.71}
\editor{DV}{violet}
\editor{AF}{Cyan}
\editor{CC}{Red}
\editor{DAL}{Green}
\editor{EM}{orange}
\editor{CH}{BluBondi} 
\editor{auth}{Red}

\begin{document}

\title{Graphene decoupling through oxygen intercalation on Gr/Co and Gr/Co/Ir interfaces}
\author{Dario A. Leon$^{1,2}$}
\email{darioalejandro.leonvalido@cnr.nano.it}
\author{Andrea Ferretti$^2$}
\author{Daniele Varsano$^{2}$}
\author{Elisa Molinari$^{1,2}$}
\author{Claudia Cardoso$^{2}$}
\email{claudia.cardoso@nano.cnr.it }
\address{$^1$FIM Department, University of Modena and Reggio Emilia, Via Campi 213/a, Modena (Italy)}
\address{$^2$S3 Centre, Istituto Nanoscienze, CNR, Via Campi 213/a, Modena (Italy)}

\begin{abstract}
We perform a density functional theory study of the effects of oxygen adsorption on the structural and electronic properties of Gr/Co(0001) and Gr/Co/Ir(111) interfaces. In both interfaces, the graphene-Co distance increases with increasing O concentration. 
The oxygen intercalation effectively decreases the electronic interaction, preventing the hybridization of graphene states with  Co $d$-orbitals, hence (partly) restoring the typical Dirac cone of pristine graphene.  In the case of graphene/Co 1ML/Ir(111), which presents a moir\'e pattern, the interplay between the O distribution and the continuous change of the graphene-Co registry can be used to tune graphene corrugation and electronic properties.
The computed electronic properties are in very good agreement with previously reported angle resolved photoemission spectroscopy and photoemission electron microscopy measurements for Gr/Co(0001).
\end{abstract}

\maketitle

\section{Introduction}
\label{sec:intro}
%
Graphene (Gr) grown on different transition metal (TM) substrates presents distinct structural characteristics and various alterations of its electronic properties depending on the lattice mismatch and degree of hybridization 
\cite{PhysRevB.84.205431, Decker2013, Pacile2013, Varykhalov_PRX_2012,Pacile_PRB_2014,Avvisati_JPCC_2017,Avvisati_NanoLett_2018, massimi2014, Usachov_2015, Vita_PRB_2014, Coraux_JPCL_2012, Cattelan_NanoScale_2015,lodesani2019graphene, Cardoso2021, Otrokov_2018,Brede2016}.
In particular, when grown on non-commensurate surfaces, it forms a moir\'e pattern and corrugates with a magnitude that depends on the interaction with the metal underneath~\cite{Decker2013,Pacile_PRB_2014,Avvisati_JPCC_2017,Avvisati_NanoLett_2018}. The corrugated Gr sheet can then be used, e.g., as a template for the adsorption of  molecules~\cite{Avvisati_JPCC_2017,Avvisati_NanoLett_2018,Avvisati2021,Wang2020,Sierda2019,Avvisati2019},
as demonstrated e.g. by the deposition of TM-phthalocyanines on Gr/Co/Ir~\cite{Avvisati_JPCC_2017,Avvisati_NanoLett_2018}.

Considering the interfaces formed by Gr, Co and Ir, we note that the in-plane lattice parameters of Gr and Co(0001) differ by only 0.05~\AA~\cite{Vincent1967,expCoO1,Pozzo_2011,PhysRevLett.107.036101}, being in practice lattice matched. In fact, Gr grown on Co(0001) lies flat on the surface, with a small interplanar distance~\cite{Usachov_2015}. The electronic properties of Gr are significantly altered by the strong interaction with the Co surface~\cite{Varykhalov_PRX_2012,Pacile_PRB_2014,Avvisati_NanoLett_2018}. On the other hand, Gr and Ir(111) lattice parameters differ by 0.25~\AA~\cite{Pozzo_2011,PhysRevLett.107.036101,Ir} and Gr grown on Ir forms a moir\'e pattern corresponding to a 10$\times$10 Gr supercell, over a 9$\times$9 Ir supercell. The Gr-Ir distance varies slightly along the moir\'e structure, depending on the local Gr-Ir registry, however the electronic properties of Gr are just barely altered by the weak interaction with the substrate.

Although Co(0001) and  Ir(111) have a different in-plane lattice parameter, when a single Co layer is intercalated underneath Gr grown on top of an Ir(111) layer, it assumes the lattice parameter of iridium~\cite{Decker2013,Pacile_PRB_2014}. In doing so, the Gr layer has an important role in stabilizing the Co single layer on the Ir surface against, e.g., the formation of Co clusters~\cite{Pacile_PRB_2014}. With 1 Co ML, the corrugation of Gr is enhanced with respect to Gr/Ir, due to the stronger Gr-Co interaction. If a larger number of Co layers is intercalated, Co recovers its bulk structure and Gr its flat configuration~\cite{Avvisati_JPCC_2017,Avvisati_NanoLett_2018}.

In this scenario, oxygen intercalation constitutes an effective way to decouple Gr from a number of metal substrates~\cite{Voloshina_2016,Bignardi_2017,Usachov2017,Jugovac2020}. Indeed, scanning tunneling spectroscopy data (STS) and density functional theory (DFT) results have shown that a Gr layer in the Gr/O/Ru(0001) interface is electronically decoupled from the substrate and Gr-derived $\pi$-states become p-doped~\cite{Voloshina_2016}. In addition, scanning tunneling microscopy (STM) experiments have shown~\cite{Voloshina_2016} that the degree of corrugation observed upon oxygen intercalation is smaller than for the pristine Gr/Ru(00001).
Likewise oxygen intercalation decouples effectively Gr from Ni(111)~\cite{Bignardi_2017}, quenching the strong substrate-adsorbate interaction with the formation of a thin Ni oxide layer at the interface.
Angle-resolved photoemission spectroscopy (ARPES) measurements on pristine Gr/Ni(111) show that the Gr $\pi$-band is shifted towards higher binding energy by about 2.5~eV with respect to the position of free-standing Gr, and the Dirac cone dissolves into the metal $3d$ bands. Upon oxygen intercalation the hybridization with Ni $3d$ states is removed and the characteristic shape of the Dirac cone is restored and identifiable near the Fermi level~\cite{Bignardi_2017}. 

Regarding Gr/Co, one of the systems under consideration in this work, recent studies based on X-ray photoemission spectroscopy (XPS), photoemission electron microscopy (PEEM) and ARPES measurements on graphene epitaxialy grown on Co(0001)~\cite{Usachov2017,Jugovac2020} show that, upon O intercalation, there is an effective electronic decoupling between Gr and Co. In fact, for an O coverage of 0.5~ML the Gr C$_{1s}$ peak shifts by 1.1~eV to lower binding energies and valence band mapping reveals that the Gr band structure acquires a nearly free-standing character with a small $p$-doping. Interestingly, it has also been shown that the O adsorption can be reversed upon annealing~\cite{Jugovac2020}.

When Co is exposed to oxygen both physical adsorption and oxidative processes may occur~\cite{exp_CoIr2,mainR}. 
While low exposures result in chemisorbed oxygen coupled ferromagnetically  with the substrate~\cite{Forster02,Getzlaff95}, when the oxygen exposure increases, cobalt oxide may form in two different stoichiometries: CoO or, in smaller proportion, Co$_3$O$_4$. A theoretical study~\cite{mainR}, investigating the surface and subsurface oxygen adsorption on Co(0001) over a wide coverage range from 0.11 to 2.0~ML, shows that the coverage with the highest adsorption energies is 0.25~ML, whereby the O atoms are adsorbed on the surface. For larger exposures, O begins to penetrate into the surface. In the same study, the energetics of O adsorption and the structural and electronic properties of the surface are discussed in detail. 
Similarly, Gr@Co interfaces can adsorb oxygen under Gr in a stable way ~\cite{Usachov2017,Jugovac2020,exp_CoIr2,exp_CoIr1}.
However, the mechanisms and structural details of O adsorption on Co depend sensitively upon the experimental parameters~\cite{exp_CoIr2,exp_CoIr1}, on the epitaxial relation between graphene and the substrate, and on the concentration of holes (carbidic islands) in the graphene layer~\cite{Jugovac2020}. In the case of epitaxially oriented graphene, the holes in the layer act as intercalation centers. 

The aim of the present work is to investigate, by means of first principles calculations, how the Gr-Co hybridization, and therefore, graphene electronic properties, change due to oxygen intercalation and, in the case of of a moir\'e structure, how the graphene corrugation pattern is influenced by the intercalation of oxygen.
We start by discussing the O adsorption on Co(0001) and Co/Ir(111) surfaces and then we consider the effects of O intercalation under graphene in both Gr/Co(0001) and Gr/1 ML Co/Ir(111) interfaces, focusing on the Gr-Co hybridization. We compute the band structure of Gr/Co(0001) before and after O adsorption and compare the results with recent ARPES and PEEM measurements~\cite{Usachov2017,Jugovac2020}. For the case of Gr/1 ML Co/Ir(111) we analyze how Gr corrugation is affected by the O intercalation and its dependence on the O distribution and concentration.

The paper is organized as follows. In Sec.~\ref{sec:methods} we describe the adopted computational methodology. In Sec.~\ref{sec:energetics} we present the results regarding the energetics of the different O adsorption sites on Co and Co/Ir surfaces. In Sec.~\ref{sec:electronic_properties}, the electronic properties of the most stable systems are investigated, and finally the effect of O intercalation on the Gr/Co interaction is discussed in Sec.~\ref{sec:GrCo_coupling} for the Gr/Co and in Sec.~\ref{sec:GrCoIr_coupling} for the Gr/1ML Co/Ir interfaces.

\section{Methods} 
\label{sec:methods}
%
We have performed DFT calculations on Co(0001) and Co/Ir(111) slabs, with and without adsorbed oxygen, as well as on the Gr/O/Co(0001) and Gr/O/1 ML Co/Ir(111) interfaces. Co and Ir bulk systems were also computed with consistent parameters for completeness.
All calculations were performed using the plane wave and pseudopotential implementation of DFT provided by the Quantum ESPRESSO package~\cite{QE1,QE2}. We employed the GGA-PBE~\cite{Perdew1996} exchange-correlation functional and adopted ultrasoft (US) pseudopotentials to describe electron-ion  interactions. The kinetic energy cutoff was set to 30 Ry for the wavefunctions and 330 Ry for the charge density. 
For the Gr/O/Co(0001) and Gr/1 ML Co/Ir(111) interfaces, van der Waals interactions were taken into account through the Grimme-D3 scheme~\cite{Grimme_2010}.

The metallic substrates were modeled by considering slabs of 5 atomic layers, referred in the text as Co(0001)$_5$ and  Co$_{1}$/Ir(111)$_{4}$, with a vacuum layer of at least 10~\AA{} in order to prevent spurious interactions between the replicas. An additional 2~\AA{} of vacuum was introduced in the modeling of the surfaces in the presence of oxygen.
The slab thickness of 5 layers was validated by calculating adsorption energies considering up to 9 layers as reported in the \suppinfo.
The Brillouin zone was sampled by using $\mathbf{k}$-points meshes of $16 \times 16 \times 16$ for cobalt and iridium bulk, $16 \times 16 \times 1$ for Co(0001) and Co/Ir(111) surfaces, and $8\times8\times1$ for surfaces with adsorbed oxygen. The projected density of states for Gr/O/Co(0001) was computed with a $40\times40\times1$ $\mathbf{k}$-grid.

\begin{figure*}
 \center
 \vspace{5pt}
  \includegraphics[width=1.0\textwidth]{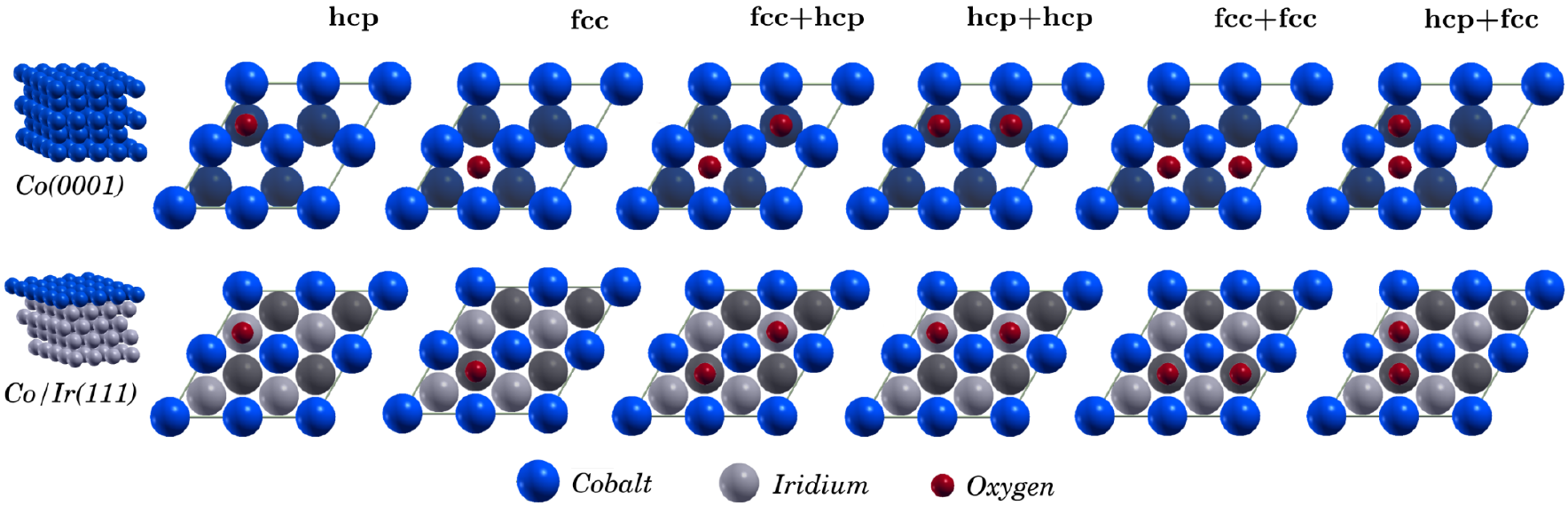}
 \caption{Top view of the different O adsorption sites considered in the calculations, for the Co (top) and Co/Ir surfaces (bottom), for 0.25 and 0.5~ML O concentrations. The labeling of the different configurations follows the notation used in Ref.~\cite{mainR}. The notation refers to surface adsorption sites. 
 }
 \label{fig:geo}
\end{figure*}

We have considered oxygen concentrations of 0.25 and 0.5~ML, that, according to Ref.~\cite{mainR}, are below the values required to form cobalt oxide and for which the oxygen atoms remain chemisorbed on the surfaces.
In order to model these oxygen coverages, we have used $p(2 \times 2)-O$ and $p(2 \times 2)-2O$ slab supercells.
The initial positions for the oxygen atoms were chosen according to the most stable configurations reported in Ref.~\cite{mainR} and are shown in Fig.~\ref{fig:geo}, where we have also labeled the O sites following the notation of Ref.~\cite{mainR}. All the atomic positions (O, Co and Ir) were then optimized until the forces acting on atoms were smaller than $1.0\times 10^{-4}$ Ry/Bohr.

The Gr/1 ML Co/Ir(111) interface was simulated considering the complete moir\'e-induced periodicity by using a 9$\times$9 supercell of Ir(111), corresponding to a 10$\times$10 supercell of pristine Gr, adding up to a supercell with more than 600 atoms. The lattice parameters were obtained by relaxing  Ir bulk at the same level of theory, corresponding in 9$\times$9 periodicity the to a hexagonal cell of 46.54 Bohr radius for the moir\'e structure. The metal slab of the Gr/1 ML Co/Ir(111) interface was modeled with four metallic layers (3 Ir plus one Co layer) with an added layer of H atoms on the bottom surface. The H layer was added in order to cancel possible interaction between states on the two surfaces due to the finite thickness of the slab, and stabilize convergence, also avoiding metastable magnetic states
(more details are given in the supplementary material). 
Atomic positions were then fully relaxed, except for the two bottom Ir layers and the H saturation layer, until ionic forces were smaller than 0.001 Ry/Bohr. The Brillouin zone was sampled with a 2$\times$2 grid of $\mathbf{k}$-points.

The oxygen adsorption energies on Co(0001) and Co/Ir(111) surfaces, $E^O_{\text{Co}}$ and $E^O_{\text{Co/Ir}}$, are defined as the difference between the total energy of the target system with the adatoms and the energy of the clean surface plus half the energy of an oxygen molecule:
\begin{eqnarray}
  E^O_{\text{Co}} &=&-\frac{1}{N_{\text{O}}} \left(E_{\text{O/Co}}-E_{\text{Co}}-N_{\text{O}}  \frac{1}{2}E_{\text{O$_2$}} \right),   \\
  E^O_{\text{Co/Ir}} &=&-\frac{1}{N_{\text{O}}} \left(E_{\text{O/Co/Ir}}-E_{\text{Co/Ir}}-N_{\text{O}}\frac{1}{2} E_{\text{O$_2$}} \right),
\end{eqnarray}
where $N_{\text{O}}$ is the number of oxygen atoms in the supercell, and $E_{\text{O/Co}}$, $E_{\text{Co}}$ and $E_{\text{O$_2$}}$ are the total energy of the adsorbate+substrate system, pristine surface, and a free oxygen molecule, respectively. 
In principle, O adsorption energies could also be referred to the energy of the free oxygen atom, resulting in an overall increase of the absorption energies, without altering the trends~\cite{mainR}, of 2.87~eV.

\section{Results and discussion}
\label{sec:results}
%
\subsection{Oxygen adsorption energies on Co and Co/Ir surfaces}
\label{sec:energetics}
%
We start by discussing the adsorption of oxygen on Co and Co/Ir interfaces. The two systems most striking differences is that 1 ML Co is epitaxial on Ir(111), resulting in a stretched lattice parameter with respect to thick Co(0001) films.
We compute the adsorption energies for oxygen adsorbed on a 5 layers Co slab (Co(0001)$_5$) and  on a slab with a layer of Co on top of a 4 Ir layers (Co$_{1}$/Ir(111)$_{4}$), for both 0.25 and 0.5 ML coverage, corresponding to one or two O atoms per cell, respectively. We consider different possible adsorption sites as illustrated in Fig.~\ref{fig:geo}, whose calculated adsorption energies are reported in Table.~\ref{tab:adsorption_energies}.

For the lowest oxygen coverage on Co(0001)$_5$, 0.25~ML, the two adsorption sites, \textit{fcc} and \textit{hcp} present very similar adsorption energies, 2.56 and 2.60~eV, respectively.
The adsorption energy drops to values around 2.13~eV for the 0.5~ML coverage, indicating a repulsive interaction between O adatoms. The most favorable adsorption configuration is the one labeled \textit{hcp}+\textit{hcp}, differing by less than 0.01~eV from the \textit{fcc}+\textit{hcp} configuration (see Fig.~\ref{fig:geo}), although such a small energy difference is at the verge of the accuracy of our computational approach, such that  both sites can be considered energetically equivalent (more details in the \suppinfo). 
Overall, the present results compare well with the values previously reported in Ref.~\cite{mainR}, with both the adsorption energy and the energy differences between different configurations decreasing when increasing O concentration. 

In the case of Co$_{1}$/Ir(111)$_{4}$, for 0.25~ML, the configuration with the lowest energy corresponds to the O atom adsorbed on a \textit{fcc} site. The adsorption energy is 2.89 eV, 0.04~eV larger then for the \textit{hcp} configuration. As in the case of the Co surface, the adsorption energy decreases with increasing O content. The most favorable configuration is then the \textit{fcc}+\textit{fcc}, followed by the \textit{fcc}+\textit{hcp} configuration, with an absorption energy of 2.64 and 2.55~eV, respectively. 
In general, the most favorable site for oxygen adsorption on Co(0001) slabs is \textit{hcp}, whereas oxygen tends to adsorb on the \textit{fcc} site for Co/Ir.  
Comparing the two surfaces, we find that the adsorption energies of oxygen for both 0.25~ML and 0.5~ML are respectively 0.29~eV and 0.45~eV larger on Co/Ir than on Co.

\begin{table}
\begin{ruledtabular}
\begin{tabular}{llcc}

\\[-4pt]
{\bf O coverage} &{\bf O site}                      & {\bf Co(0001)$_5$}        & {\bf Co$_{1}$/Ir(111)$_{4}$} \\
\\[-4pt]
\hline\\[-3pt]
{ 0.25 ML}       & \textit{hcp}                      & 2.60 [2.67]	            & 2.85\\
                 & \textit{fcc}                      & 2.56 [2.66]              & 2.89\\
\\[-4pt]
\hline\\[-3pt]
{ 0.5 ML}        & \textit{fcc}+\textit{hcp}         & 2.18 [2.13]         & 2.55	               \\
                 & \textit{hcp}+\textit{hcp}         & 2.19 [2.09]         & 2.53	               \\
                 & \textit{fcc}+\textit{fcc}         & 2.13 [1.97]         & 2.64	               \\
                 & \textit{hcp}+\textit{fcc}         & 1.78 [{ }{ - }{ }]  & 2.22                  \\[3pt]
\bottomrule
 \end{tabular}
 \end{ruledtabular}
  \caption{Adsorption energies, in eV/O, computed for Co(0001)$_5$ and Co$_{1}$/Ir(111)$_{4}$. For the sake of comparison, we show in brackets the results from Ref.~\cite{mainR}. 
  }
 \label{tab:adsorption_energies}
\end{table}

The computed distances between the oxygen adatoms and the surface are reported in Table~\ref{tab:OCoIr_struct} as $d_O$.
For the 0.25~ML configurations, the higher the adsorption energy the closer the oxygen atoms are to the surface. The two configurations of adsorbed O on Co(0001) show similar O-surface distances, that differ in less than 0.06~\AA, while on Co/Ir the  differences are larger, 0.14~\AA.
%
For 0.5 ML, the O-surface distances, for both Co(0001)$_5$ and Co$_{1}$/Ir(111)$_{4}$, have a larger dependence on the absorption site, with no apparent correlation between the O-surface distances and the adsorption energies, probably due to the O-O interaction. 
%
Comparing the lowest energy configurations of each surface, the oxygen atoms are closer to the surface on Co/Ir than on Co, consistently with its larger in-plane lattice parameter.

Oxygen adsorption affects also the layer-layer distance in the metallic slabs, as shown in Tab.~\ref{tab:OCoIr_struct}, with an increase of the first interlayer distance, $d_{z1}$, on both Co(0001)$_5$ and Co$_{1}$/Ir(111)$_{4}$, with the O concentration.  
In contrast, the distance between the second and third layers, $d_{z2}$, presents a slightly decrease.

\begin{table}
\begin{ruledtabular}
\begin{tabular}{llcccccc}
\\[-7pt]
{\bf O coverage} & {\bf O site}  & \multicolumn{3}{c}{{\bf Co(0001)$_5$}}        & \multicolumn{3}{c}{{\bf Co$_{1}$/Ir(111)$_{4}$}} \\[2pt]
  &   & $d_O$       & $d_{z1}$   & $d_{z2}$              & $d_O$       & $d_{z1}$   & $d_{z2}$\\[2pt]
\hline\\[-5pt]
 pristine     &-&  -             & 3.69          & 3.83             &  -            & 3.82          & 4.33 \\[2pt] 
\hline\\[-5pt]
 0.25~ML
&   \textit{hcp}             & 2.12         & 3.75          & 3.80             & 1.98          & 3.89          & 4.31 \\
&   \textit{fcc}            & 2.18          & 3.78          & 3.78             & 1.84          & 3.90          & 4.30 \\[2pt]
\hline\\[-5pt]
 0.5~ML
&  \textit{fcc}+\textit{hcp}         & 2.24          & 3.86          & 3.76             & 1.92          & 3.95          & 4.29 \\
&   \textit{hcp}+\textit{hcp}        & 2.07          & 3.82          & 3.78             & 1.90          & 3.94          & 4.30 \\
&   \textit{fcc}+\textit{fcc}        & 2.06          & 3.87          & 3.75             & 1.80          & 3.97          & 4.29 \\
&   \textit{hcp}+\textit{fcc}        & 2.22          & 3.90          & 3.78             & 1.88          & 3.97          & 4.29 \\[2pt]
\bottomrule
\end{tabular}
\end{ruledtabular}
 \caption{Oxygen-substrate ($d_0$) and interlayer distances ($d_{z1}$,$d_{z2}$) for Co(0001)$_5$ and Co/Ir(111)$_4$. For the Co(0001)$_5$ surface, the interlayer distances correspond to the distances between Co neighboring layers. For the Co$_{1}$/Ir(111)$_{4}$ surfaces, the first interlayer distance $d_{z1}$ corresponds to the distance between the Co layer and the next Ir layer, while the second one ($d_{z2}$) corresponds to an Ir-Ir layer distance. All the reported distances are in \AA\ and were computed considering the difference between the average vertical coordinate of each atomic layer.  
 }
  \label{tab:OCoIr_struct}
 \end{table} 

We have also studied configurations with the oxygen atoms inside the metal surface. For both Co(0001)$_5$ and Co$_{1}$/Ir(111)$_{4}$, and for both O concentrations considered, the added O always migrates to the surface. The one exception is a configuration with O occupying an \textit{fcc} site below the second metal layer. However, this subsurface configuration has an adsorption energy that is considerably smaller than the surface configurations.  In fact, subsurface O adsorption is expected to occur only for O contents larger than 1~ML~\cite{mainR}.
In Ref.~\cite{mainR} oxygen is found below the first layer, but this contrast may be justified e.g. by differences in the relaxation process. We report more details about the subsurface configurations in the \suppinfo.

\subsection{Electronic and magnetic properties}
\label{sec:electronic_properties}
%
\begin{figure}
  \centering
  \includegraphics[width=0.48\textwidth]{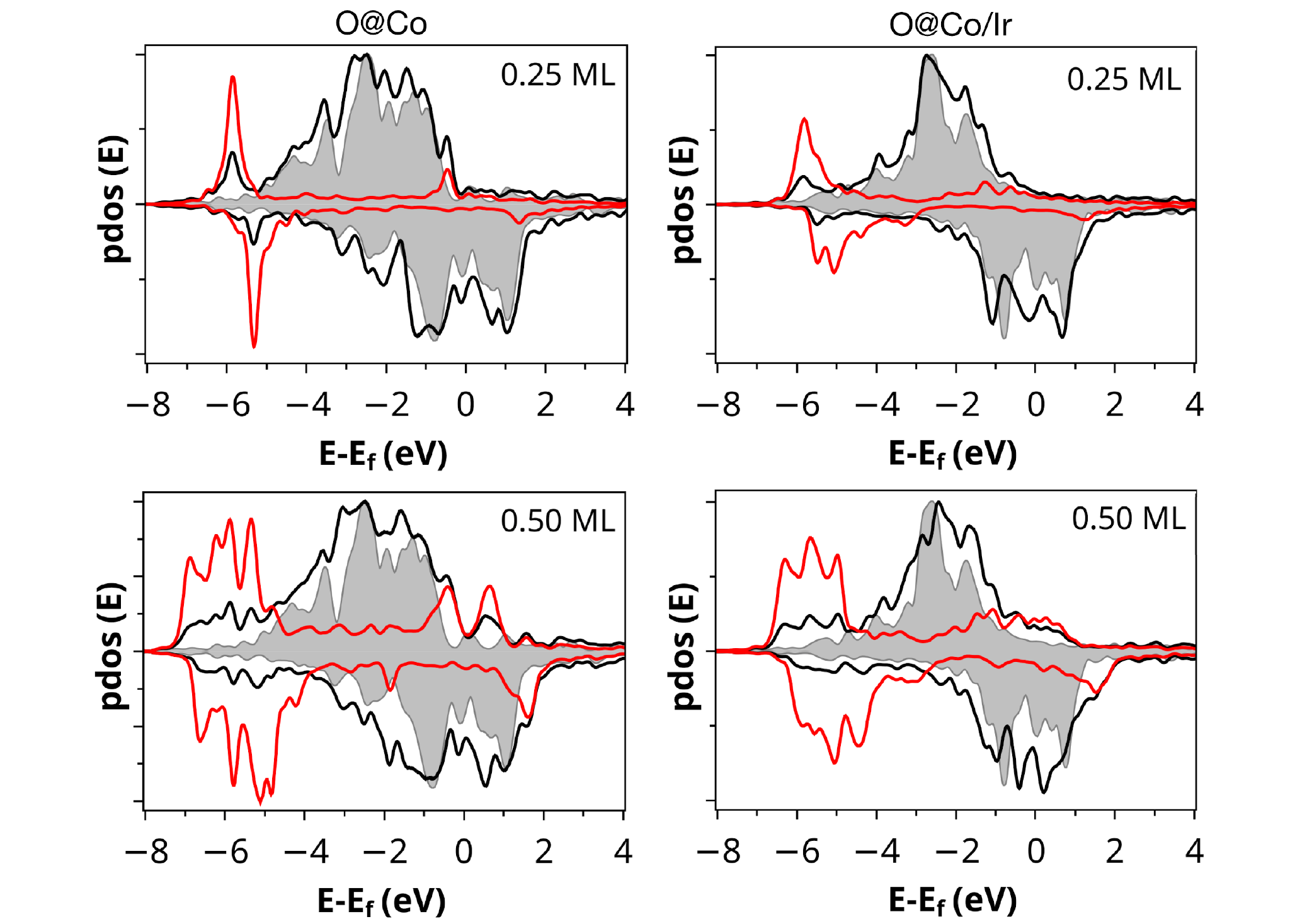}
 \caption{Density of states computed for O@Co (left panels) and O@Co/Ir (right panels), for 0.25 (top panels) and 0.5 ML (bottom panels). The density of states is projected on the atomic orbitals of O (red lines) and Co on the top layer (black lines). The Co top layer DOS for the pristine surface is shown is grey.}
 \label{fig:DOS}
\end{figure}

In order to understand the influence of O on the Co electronic properties, in Fig.~\ref{fig:DOS} we report the density of states projected on the O and the top layer Co atoms (pDOS). Majority and minority spin bands are shown for  0.25 and 0.5 ML coverage, for both Co(0001)$_5$ and Co$_{1}$/Ir(111)$_{4}$, and compared with the pristine systems (displayed in grey). 
The DOS computed for the clean systems show the Co $3d$ states localized around the Fermi energy ($E_F$), with a spin split of about 2~eV, that results in a magnetic moment of $1.68~\mu_B/\text{Co}$ for Co(0001)$_5$ and  $2.02~\mu_B/\text{Co}$ for Co$_{1}$/Ir(111)$_{4}$. 
The majority spin $d$-states are completely occupied, whereas the minority DOS cuts through the Fermi level, with peaks above and below, and a valley at Fermi for both systems.

The Co$_{1}$/Ir(111)$_{4}$ shows a narrower DOS than Co(0001)$_5$, leading to larger magnetic moments. This enhancement is related to some degree of confinement induced when going from a Co slab to a single layer on Ir~\cite{Avvisati_NanoLett_2018} as well as to the stretched lattice parameter of Co 1ML due to the presence of Ir(111).
The same narrowing of the DOS leading to large magnetic moments was reported also for the Gr/Fe/Ir~\cite{Cardoso2021} and Gr/FeCo/Ir~\cite{Pacile2021} interfaces, both from experiment and DFT calculations. It is worth noticing, though, that these calculations were performed within DFT+U~\cite{cococcioni2005linear}  and that the choice for the Hubbard $U$ parameter affects the computed magnetic moments, as described in detail in the Supplementary material of Ref.~\cite{Pacile2021}.

In the presence of O adatoms there is a clear overlap between Co and O states, that induces a spin-split of the latter. The O $2p$ states hybridize with the lowest edge of the Co $3d$ bands, resulting in a peak around 5.5~eV below the Fermi level. There is a second state, with the majority spin located just below the $E_F$ and the minority spin about 1.5~eV above in the case of the 0.25 ML, more pronounced for Co(0001)$_5$ rather than Co$_{1}$/Ir(111)$_{4}$. 
For the 0.5 ML, the O pDOS becomes wider and shifts slightly downward (i.e. towards larger binding energies) reaching to 6.8~eV, due to the O-O repulsive interaction, in agreement with Ref.~\cite{mainR}.
Comparing the Co(0001)$_5$ and Co$_{1}$/Ir(111)$_{4}$ surfaces, there are evidences of a larger hybridization of the O atom with Co states in the case of Co$_{1}$/Ir(111)$_{4}$, for which the O pDOS peaks are broader, consistently with the larger O adsorption energies mentioned above.

\begin{table}
\begin{ruledtabular}
\begin{tabular}{llcccccc}
\\[-7pt]
{\bf O coverage} &   {\bf O site}    & \multicolumn{3}{c}{{\bf Co(0001)$_5$}}                          & \multicolumn{3}{c}{{\bf Co$_{1}$/Ir(111)$_{4}$}} \\[2pt] 

 &   & $m_\text{tot}$    & $m_\text{Co}$  & $m_\text{O}$      & $m_\text{tot}$  & $m_\text{Co}$    & $m_\text{O}$    \\[2pt]
\hline\\[-5pt]
pristine   &              & 1.68          & 1.76     & -            & 2.02       & 1.92        & -          \\[2pt] 
\hline\\[-5pt]
0.25~ML    &
   \textit{hcp}                    & 1.68          & 1.69     & 0.23         & 2.07       & 1.89        & 0.37      \\
&   \textit{fcc}                   & 1.70          & 1.72     & 0.26         & 2.06       & 1.85        & 0.31      \\[2pt]
\hline\\[-5pt]
 0.5~ML    &   \textit{fcc}+\textit{hcp}    & 1.61          & 1.24     & 0.08         & 2.02       & 1.73       & 0.37       \\
           &   \textit{hcp}+\textit{hcp}    & 1.67          & 1.49     & 0.18         & 2.12       & 1.80       & 0.33        \\
           &   \textit{fcc}+\textit{fcc}    & 1.68          & 1.53     & 0.19         & 1.94       & 1.68       & 0.26       \\
           &   \textit{hcp}+\textit{fcc}    & 1.71          & 1.64     & 0.27         & 2.29       & 1.93       & 0.43       \\[2pt]
 \end{tabular}
\end{ruledtabular}
 \caption{Total magnetization (in units of Bohr magneton) per cobalt atom ($m_\text{tot}$, in $\mu_B$/Co), average magnetic moment for the Co atoms in the first layer ($m_\text{Co}$),  and average O magnetic moment ($m_\text{O}$), computed for O adsorbed on Co(0001)$_5$ and Co$_{1}$/Ir(111)$_{4}$, for both the 0.25 and 0.5 ML coverage. 
 }
  \label{tab:OCoIr_mag}
\end{table} 

In Table~\ref{tab:OCoIr_mag} we show a summary of the magnetization and magnetic moments obtained for O adsorbed on the Co(0001)$_5$ and Co$_{1}$/Ir(111)$_{4}$ slabs. In all configurations the O moment is aligned ferromagnetically with Co. 
For the pristine systems, the average magnetic moment of the top layer Co atoms is larger for Co$_{1}$/Ir(111)$_{4}$  than for the Co(0001)$_5$ slab, 1.92 and 1.76~$\mu_B$ respectively, reflecting the narrower structure of the DOS observed for Co$_{1}$/Ir(111)$_{4}$. After oxygen adsorption, a small magnetic moment is found on the O atom, with values ranging from 0.08 to 0.27~$\mu_B$ on Co(0001)$_5$, and from 0.26 to 0.43~$\mu_B$, on Co$_{1}$/Ir(111)$_{4}$, depending on the oxygen concentration.
However, chemisorbed oxygen atoms have a very small impact on the total magnetization of the Co and Co/Ir slabs since the moments of the Co atoms in the first layer, ($m_{\text{Co}}$ in Table~\ref{tab:OCoIr_mag}), tend to decrease with respect to the pristine systems.
The effect is slightly larger for Co$_1$/Ir(111)$_4$ since it has only one magnetic layer (i.e. the top Co layer).
Among the most energetically stable configurations, the magnetic moment varies less than 5\%. 

Cobalt oxide bulk (CoO) is known to have an antiferromagnetic (AF) order, that has been well-described at the DFT+U level e.g. in Refs.~\cite{presCoO,dopCoO,Anisimov91,Tran06,Wdowik07}. This motivated the search for possible antiferromagnetic configurations also for the surfaces studied in this work. In this case, the only AF solutions were found for two Co/Ir configurations, with O adsorbed on \textit{hcp}+\textit{hcp} and \textit{fcc}+\textit{fcc}. However these configurations have a total energy that is 0.3 eV/O higher than the ferromagnetic configuration, being thereby significantly less stable. More details are provided in the \suppinfo.

\subsection{Graphene on Co: decoupling through O intercalation}
\label{sec:GrCo_coupling}
%
\begin{figure}
  \centering
   \includegraphics[width=0.45\textwidth]{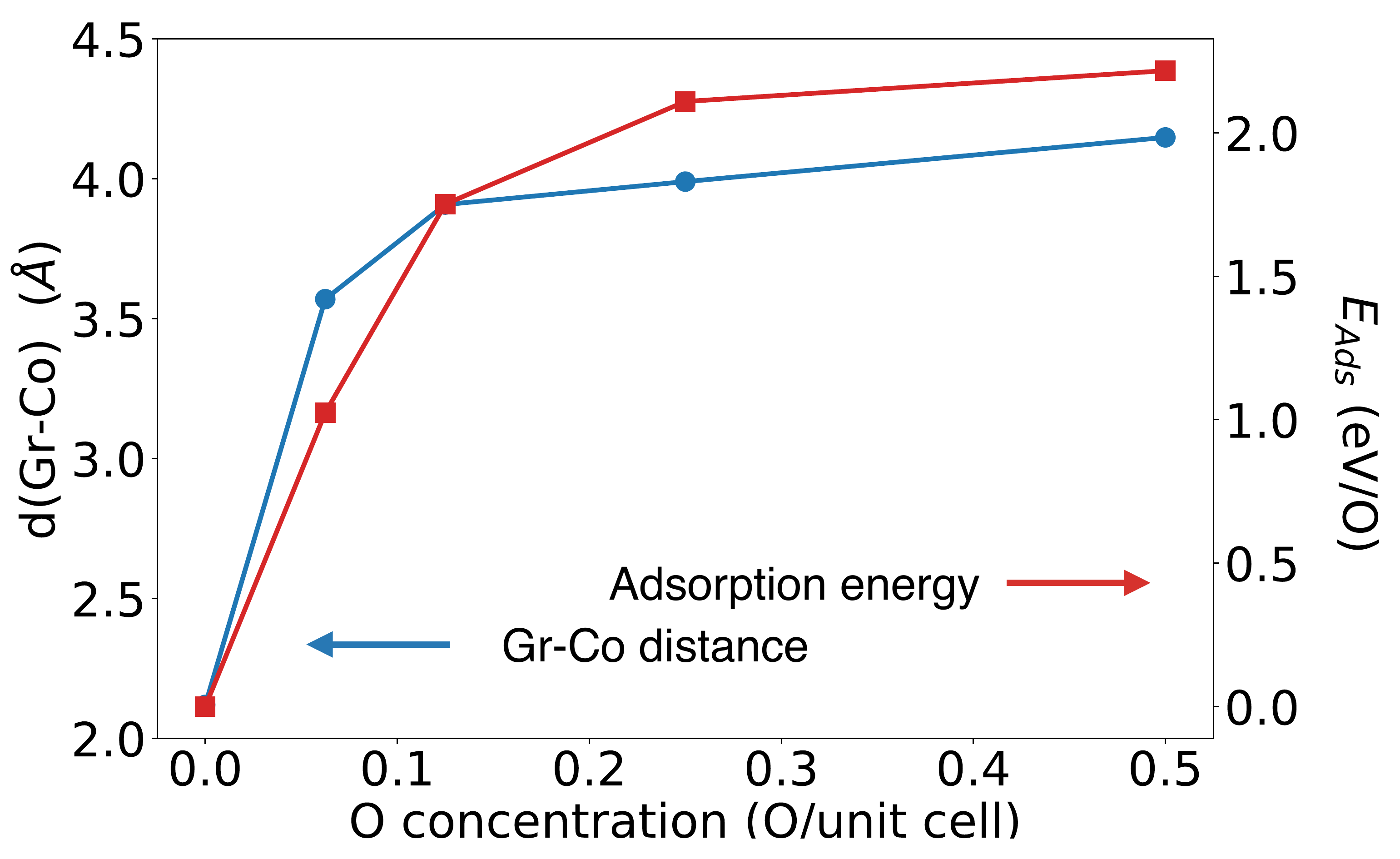}
 \caption{Gr-Co distance (blue line, left scale) and corresponding adsorption energy (red line, right scale), computed for different O concentrations.}
 \label{distance_vs_O}
\end{figure}
\begin{figure*}
  \centering
   \includegraphics[width=0.85\textwidth]{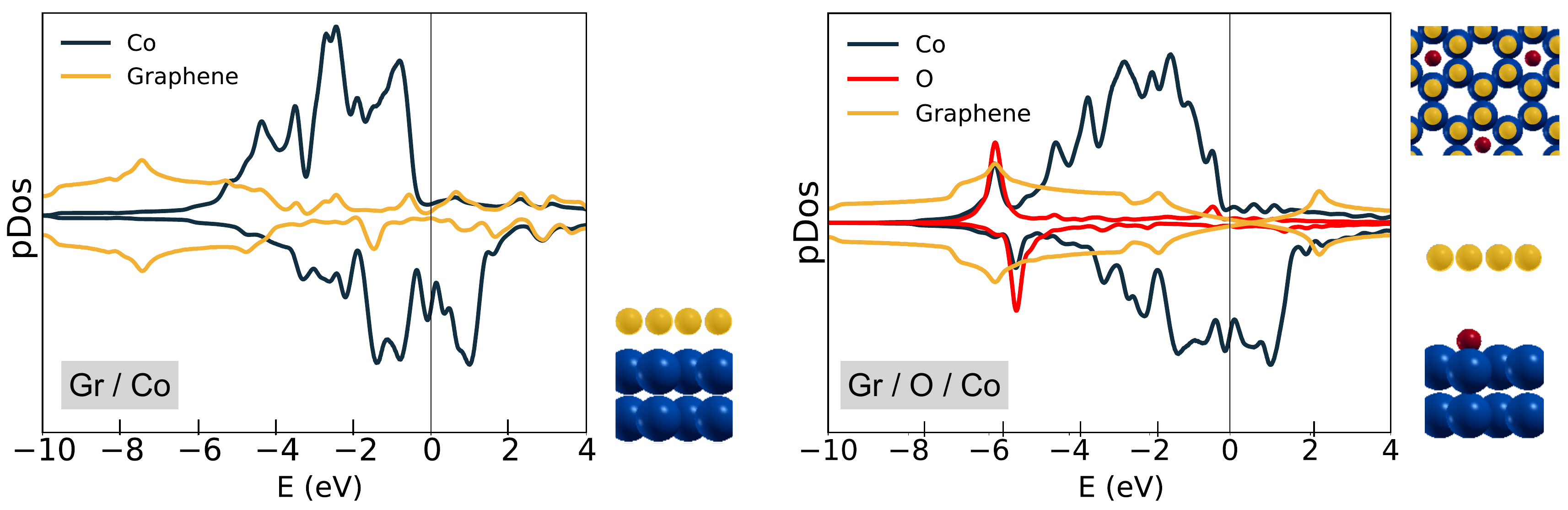}
 \caption{Density of states computed for Gr@Co (left panel) and Gr@O/Co (right panel) and projected on the C, on the first layer Co and O atomic orbitals. Next to each plot, a scheme with the atomic structure, illustrating the difference in the Gr-Co distance for the two systems and, for the case of Gr@O/Co, a top view showing the O distribution. }
 \label{dos_GrOCo}
\end{figure*}

Having established structural, electronic and magnetic properties of oxygen adsorbed on Co and Co/Ir, now we consider the case of graphene (Gr) adsorbed on Co(0001) and Co/Ir(111), in the presence of O intercalation.
In particular, we take advantage of the structural analysis performed in Sec.~\ref{sec:energetics} to build the models for Gr/O/Co(0001) and Gr/O/Co/Ir(111). XPS, PEEM and ARPES measurements on Gr/Co(0001) show that, upon O intercalation, there is an electronic decoupling between graphene and Co, and that the graphene band structure acquires a nearly free-standing character except for a small p-doping~\cite{Usachov2017,Jugovac2020}.
In order to assess the effect of oxygen intercalation under Gr, we have performed calculations for graphene on Co(0001)$_5$, with and without O adsorbed under graphene. We consider an oxygen coverage ranging from 0.0625 to 0.5 ML.

In Fig.~\ref{distance_vs_O} we show, in red, the adsorption energy, which increases for increasing O concentration. For 0.25 and 0.5~ML the adsorption energies are 2.11 and 2.21~eV respectively, similar to the values computed without the graphene layer (2.60 and 2.19~eV). However, without graphene the adsorption energy for 0.25~ML is slightly larger than the one for 0.5~ML, which indicates that the graphene layer partially screens the interaction between neighboring oxygen atoms.

Concerning structural properties, in Fig.~\ref{distance_vs_O} we show the change of the average distance between graphene and the Co surface. We observe that oxygen adsorption has a large impact on the structure of the Gr/Co interface yielding a dramatic increase in the graphene-Co distance, that changes from 2.1~\AA{} in the pristine case to 4.1~\AA{} upon the adsorption of 0.5 ML of O. Moreover, even the smallest O concentrations considered here lead to a significant increase on the Gr-Co distance.
In Fig.~\ref{dos_GrOCo} we also present the density of states computed for the pristine GrCo interface and upon O adsortion, again with a 0.25~ML concentration.
In the absence of oxygen, a clear hybridization between Co $d$ electrons and C states close to the Fermi level can be seen from the projected density of states of Gr@Co, as shown in the left panel of Fig.~\ref{dos_GrOCo}, and as already observed in Refs.~\cite{Varykhalov_PRX_2012,Pacile_PRB_2014,Calloni_2020}. In the presence of adsorbed oxygen, the  {pDOS} of C atomic orbitals is similar to that of freestanding graphene. 

\begin{figure}
\center
\includegraphics[width=0.47\textwidth]{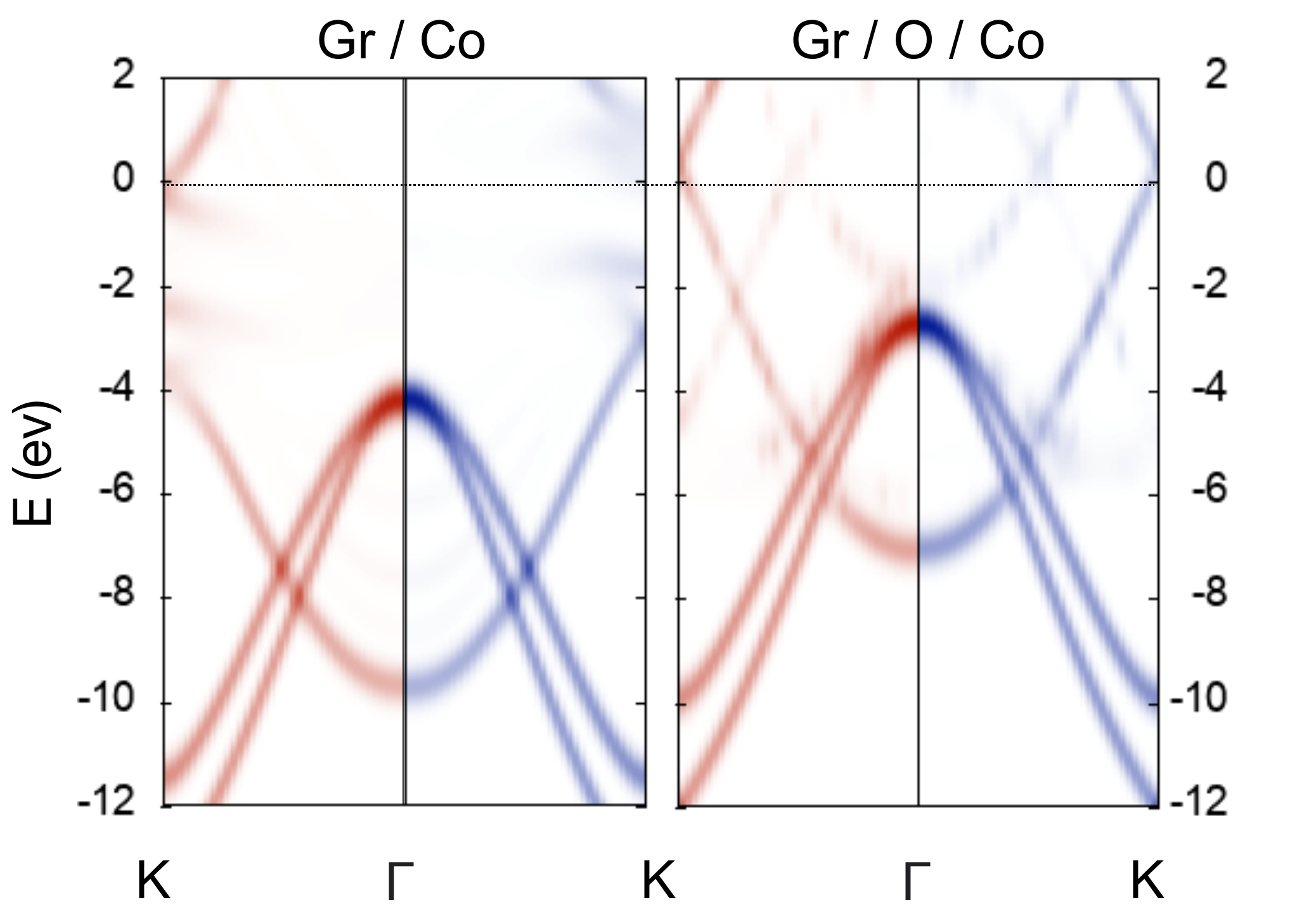}
 \caption{Bands projected on the atomic orbital of graphene, computed for Gr@Co (left panel) and Gr@OCo (right panel) unfolded in the graphene unit cell. The red and blue colors correspond to the spin up and spin down states.}
 \label{bands_GrOCo}
\end{figure}

In Fig.~\ref{bands_GrOCo} (left panel) the bands computed for the 2$\times$2 Gr@Co unit cell are mapped into the Gr 1$\times$1 Brillouin zone by using the \texttt{unfold-x} code~\cite{unfold-x}.
This procedure allows us to visualize an effective band structure corresponding to the graphene unit cell in the presence of substrates and adatoms.
Similarly to what found in Refs.~\cite{Cardoso2021,Pacile2021} for the GrCoIr interface in a 10$\times$10 Gr supercell, the Dirac cone is disrupted by the hybridization of Gr with the Co $d$ states, and the C states show a spin-split induced by the spin-polarized Co states. Overall, this results into a shift of the apex of the cone of about -3.7 and -2.9~eV for the majority and minority spin channels, respectively (left panel of Fig.~\ref{bands_GrOCo}). 

At variance with Gr@Co, for Gr@O/Co (right panel of Fig.~\ref{bands_GrOCo}) the Dirac point can be recognized slightly above the Fermi level, at about 0.4~eV, indicating a small doping of graphene, in excellent agreement with the value of 0.3~eV recently measured by ARPES experiments~\cite{Usachov2017}, and the 0.4~eV value from the PEEM momentum map reported in Ref.~\cite{Jugovac2020}. The same type of doping is also seen in Gr@Ir~\cite{Calloni_2020,Scardamaglia_JPCC_2013} and in Gr/Ru(0001) upon O intercalation~\cite{Voloshina_2016}, with the Dirac cones located at +0.21 and +0.5~eV, respectively.

The oxygen adsorption also changes the magnetic moments of the interface. Without O, the average magnetic moment of Co is 1.7~$\mu_B$, but the hybridization with graphene decreases the moment of the Co top layer. In fact, the average magnetic moment increases from 1.6 to 1.8~$\mu_B$ when going from the top to the fourth Co layer, in good agreement with the values estimated from X-ray magnetic circular dichroism, $\mu_s=1.47 \mu_B$~\cite{Jugovac2020}. Concerning graphene, there is a small magnetic moment induced in the two nonequivalent C atoms, -0.08 and 0.05~$\mu_B$. Upon O adsorption the Co magnetic moment is more uniform across the layers, with an average of 1.7~$\mu_B$, and there is no significant moment on graphene. The adsorbed O atom has a magnetic moment of 0.2~$\mu_B$, in the same range of the ones computed without the graphene layer.

\begin{figure*}
\center
\includegraphics[width=0.8\textwidth]{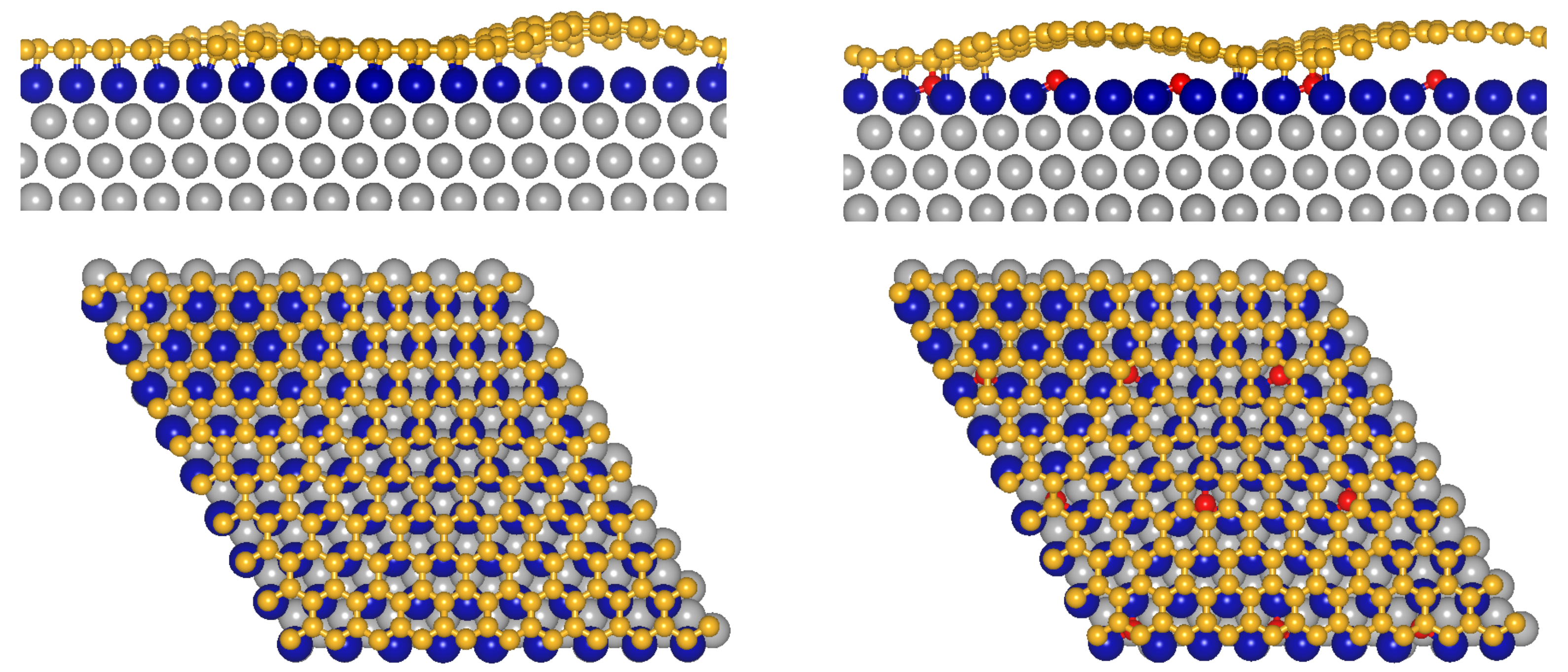}
 \caption{Scheme of the crystal structure as computed within DFT for Gr/1ML Co/Ir before and upon O adsorption, with a O concentration of $11\%$.}
 \label{structure_GrOCoIr}
\end{figure*}

\begin{figure*}
\center
\includegraphics[width=1.0\textwidth]{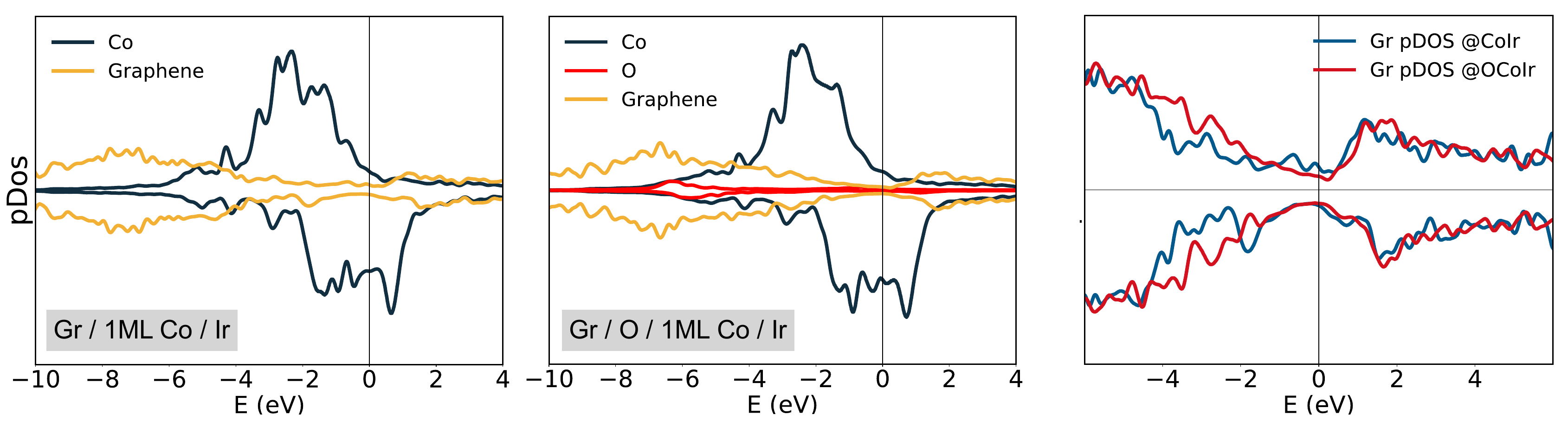}
 \caption{Density of states computed for Gr/1ML Co/Ir, without O (first plot) and with an O concentration of $11\%$ (second plot) and projected on the Co, C and O atomic orbitals. The third plot compare the Gr DOS of the two systems.}
 \label{dos_GrOCoIr}
\end{figure*}

\subsection{Graphene on CoIr: interplay between O and GrCo registry}
\label{sec:GrCoIr_coupling}

We will now focus on the effects of O intercalation on the Gr/1 ML Co/Ir(111) interface. As mentioned before this interface presents a moir\'e structure corresponding to a 10$\times$10 Gr supercell on top of a 9$\times$9 Ir supercell, intercalated by 1 ML Co with a structure similar to Ir, as illustrated in Fig.~\ref{structure_GrOCoIr}. The graphene sheet presents a corrugation caused by the fact that the Gr-Co distance depends on the registry between the two layers~\cite{Pacile_PRB_2014}, which varies continuously through the moir\'e unit cell. According to the present results, when Gr sits on top of a Co atom, the Gr-Co distance is minimum, with a value of 1.94~\AA\ and the largest Gr-Co distance is of 3.5~\AA\ and corresponds to hollow sites (\textit{fcc}/\textit{hcp} stacking). This results in a corrugation of the Gr sheet of 1.59~\AA, in good agreement with the literature~\cite{Decker2013,Avvisati_NanoLett_2018,Avvisati_JPCC_2017,Duncan2019}.

We performed a similar geometry optimization of the interface before and after O adsorption. We have considered an uniform distribution of O in the supercell corresponding to a concentration of $11\%$. The computed adsorption energy is 1.17~eV/O, slightly smaller than the Gr/Co(0001) value interpolated for the same O concentration, 1.6~eV/O.
The optimized structures are illustrated in Fig.~\ref{structure_GrOCoIr}. The comparison of the two interfaces shows, as for the case of the Gr/Co(0001) interface, that O causes an increase of the average Gr-Co distance from 2.24~\AA{} in the pristine case to 2.96~\AA{} upon O intercalation, and a change on the structure of Gr corrugation. Despite this increase, the small O concentration and the particular atomic distribution considered in the present calculations allows for some regions, corresponding to C sites on top of Co atoms, to preserve a small Gr-Co distance, which also results in a slight increase in the corrugation, from 1.59 \AA{} in the pristine case to 1.74~\AA. 

This would depend on the O distribution on the surface and we expect that, for large enough O concentrations the Gr layer completely decouples from the Co surface, regaining its free standing properties, as seen in the previous section for the Gr/Co(0001) interface. In Fig.~\ref{dos_GrOCoIr} we show the projected density of states corresponding to these geometries. Comparing with the DOS obtained with and without O intercalation, shown in the left panel of Fig.~\ref{dos_GrOCo}, a small O concentration already results in a partial recover of the Gr free standing character, with a decrease of its spin polarization, even if there are still signs of Gr-Co hybridization.

\section{Conclusions}
%

In this work we study, by means of DFT calculations, the effects of oxygen intercalation on the electronic properties of the Gr/Co(0001) and Gr/Co/Ir(111) interfaces.
In the first part of the work we disregard the presence of Graphene and
we  study two different surfaces in the presence of oxygen adatoms: Co(0001) and a  single layer of Co deposited on of Ir(111), taking into account that in the latter case the single Co layer is epitaxial on Ir(111)~\cite{Decker2013,Pacile_PRB_2014}, and therefore has a lattice parameter 0.25~\AA\ larger than that of the Co(0001) surface.
In particular, this allows us to address the effect of the strain on the absorption of oxygen on Co.
The adsorption energies are about 0.3~eV larger for Co/Ir(111) than for the Co(0001) surface and, in both structures, the adsorption energies decrease with increasing O content. 

Regarding the electronic properties, upon adsorption there is a clear hybridization between O and Co states, that is larger in the case of Co/Ir(111). The larger magnetic moment of the Co single layer when deposited on Ir, a consequence of its strained lattice, also results in a slightly larger O magnetic moment.

Finally, we have addressed the effect of O intercalation in the Gr/Co(0001) and Gr/Co/Ir(111) interfaces. The present results show that O adsorption induces an increased graphene-Co distance, and effectively reduces the electronic interaction with the Gr layer. The graphene Dirac cone, disrupted by the hybridization with the Co states around the Fermi energy, is restored upon O intercalation. The graphene bands recover their free-standing character except for a small p-doping of graphene since its Dirac cone is now located slightly above the Fermi energy, in excellent agreement with ARPES~\cite{Usachov2017} and PEEM~\cite{Jugovac2020} measurements reported for  Gr/Co(0001). In the case of Gr/Co 1ML/Ir(111), which presents a moir\'e pattern, the presence of the O atoms changes locally the Gr-Co distance. 

We have shown that, playing with the O distribution and the continuous change of the graphene-Co registry, it is possible to tune both graphene corrugation and its electronic properties. These results support the idea that O intercalation on Gr/Co interfaces, shown experimentally to be a reversible process~\cite{Jugovac2020}, could be used to tune the Gr-Co structural and electronic properties, which depend on both on the O concentration and surface distribution.

\section*{Acknowledgments}
%
We acknowledge stimulating discussions with M.G. Betti, C. Mariani and G. Avvisati.  This work was partially supported by the MaX -- MAterials design at the eXascale -- European Centre of Excellence, funded by the European Union program H2020-INFRAEDI-2018-1 (Grant No. 824143). Computational time on the Marconi100 machine at CINECA was provided by the Italian ISCRA program.

\bibliographystyle{unsrt}
\bibliography{paper_arxiv}

\clearpage


\section*{Supplementary material}

# Graphene decoupling through oxygen intercalation on Gr/Co and Gr/Co/Ir interfaces: Supplemental material

Dario A. Leon[1,2,*], Claudia Cardoso[2], Daniele Varsano[2], Elisa Molinari[1,2], and Andrea Ferretti[2]

[1]FIM Department, University of Modena and Reggio Emilia, Via Campi 213/a, Modena (Italy) and
[2]S3 Centre, Istituto Nanoscienze, CNR, Via Campi 213/a, Modena (Italy)

## I. METHODOLOGY

### A. Validation

In order to validate the computational method described in the main text, we started by calculating the structural parameters for Co, Ir and the oxygen molecule, using both ultra-soft (US) and norm-conserving (NC) pseudopotentials. The results are summarized in Table S1, and compared with values previously reported, both experimental [1] and computed with a similar level of theory [2, 3]. The computed lattice parameter of $hcp$ cobalt bulk is $a = 2.515$ Å and $c/a = 1.617$, which differs from the experimental values by less than 0.4%. For the $fcc$ iridium bulk we obtain a lattice parameter of $a = 3.896$ Å, which differs by less than 1.5% from the experimental one.

In case of the oxygen molecule, we obtained for the bound length 1.229 Å, which differs by less than 1.7% from the experimental value, while the binding energy, 2.867 eV/O, differs by less than 12%. This typical overestimation is well-known [2, 3] and due to the sensitivity of molecular bindings on the details of the pseudopotential. All our calculations are in very good agreement with the existent literature, such small differences are comparable with the level of accuracy of DFT.

| | Exp. [2, 4] | NC [2] | US | NC |
|---|---|---|---|---|
| Co, $a$(Å) | 2.507 | 2.52 | 2.515 | 2.497 |
| Co, $c/a$ | 1.623 | 1.62 | 1.617 | 1.618 |
| Ir $a$(Å) | 3.839 | - | 3.896 | 3.870 |
| $d_{O-O}$ (Å) | 1.21 | 1.23 | 1.229 | 1.230 |
| $E_O - \frac{1}{2}E_{O_2}$ (eV) | 2.56 | 2.89 | 2.867 | 2.878 |

TABLE S1: Comparison of structural parameters and formation energy with experiments [4], and previous DFT PBE calculations [2]. Calculations using ultra-soft (US) and norm-conserving (NC) pseudopotentials done in the present work are also reported.

| O adsorption (ML) | site | Co | | | Co/Ir | |
|---|---|---|---|---|---|---|
| | | NC [2] | US | NC | US | NC |
| **0.25** | **hcp** | 5.56 | 5.47 | 5.51 | 5.72 | |
| | **fcc** | 5.55 | 5.43 | | 5.76 | 5.81 |
| **0.5** | **fcc+hcp** | 5.02 | 5.05 | 5.10 | 5.42 | |
| | **hcp+hcp** | 4.98 | 5.06 | 5.08 | 5.40 | |
| | **fcc+fcc** | 4.86 | 5.00 | | 5.51 | |
| | **hcp+fcc** | - | 4.65 | | 5.09 | |
| **0.5** | **hcp+fcc(octa)** | 2.95 | | | | |
| | **hcp+fcc(octa')** | - | 4.16 | | 3.22 | |

TABLE S2: Comparison of the adsorption energies for Co and Co@Ir obtained with ultra-soft and norm-conserving pseudopotentials.

### B. Pseudopotentials

The differences in the structural parameters that we obtained in Table S1 when using NC and US pseudopotentials are very small. Nevertheless, as reported in Table S2, we checked also the differences in the adsorption energies of few cases of O adsorbed on Co and Co@Ir respectively. The results of NC and US pseudopotentials are very consistent with each other. In the case of the 50% oxygen coverage, the $fcc + hcp$ and the $hcp + hcp$ configurations are very closed in energy but the small difference results in a different ground state when using different pseudopotentials.

### C. Periodic boundary conditions: Replica distance

The Co(0001) and Co/Ir(111) surfaces were modelled using periodic boundary conditions, considering slabs surrounded by a vacuum layer, thick enough to prevent spurious interactions between the replica. In Table S3 we present the dependence of the adsorption energy of oxygen in slabs of 5 layers of Co and Co/Ir on the vacuum thickness. The difference in the adsorption energy when using 10 to 12 Å is below 0.001 eV for all the systems, therefore we used a 12 Å vacuum layer in the calculations presented in the main text.

*Electronic address: darioalejandro.leonvalido@cnr.nano.it

| O (ML) | adsorption site | E$_{ad}$ (eV) | | |
|---|---|---|---|---|
| | | 10 Å | 12 Å | 15 Å |
| **0.25 @Co$_5$** | **hcp** | 5.470 | 5.470 | |
| | **fcc** | 5.428 | 5.429 | |
| **0.5 @Co$_5$** | **fcc+hcp** | 5.052 | 5.051 | |
| | **hcp+hcp** | 5.060 | 5.060 | |
| | **fcc+fcc** | 5.003 | 5.002 | |
| | **hcp+fcc** | 4.647 | 4.646 | |
| **0.25 @Co/Ir$_4$** | **hcp** | 5.717 | 5.717 | 5.717 |
| | **fcc** | 5.756 | 5.756 | 5.756 |

TABLE S3: Adsorption energies computed for different amounts of vacuum for O adsorbed on Co$_5$ and Co/Ir$_4$ slabs.

### D. Slab Thickness

The Co(0001) and Co/Ir(111) surfaces were modeled considering slabs with a limited number of atomic layers. In order to converge the results with respect to the slab thickness, we performed calculations of the adsorption energy with respect to the number of atomic layers. In Fig. S1 we show the values for 4 to 9 layers at 0.25 ML, the most reactive oxygen coverage, for Co and Co/Ir. When increasing the number of layers, the adsorption energy shows some fluctuations, that are of about 0.02 eV for Co/Ir and much larger for Co, about 0.1 eV. The difference is probably due to the increase of the total magnetization of the Co slab when adding a new layer. The observed fluctuations seem to follow the vertical periodicity of the Co stacks: Co has an A-B stacking, with a periodicity of two layers. For Co/Ir, an extra Ir layer does not change the magnetization and the adsorption energy does not change beyond 7 layers. Despite the fluctuations, the order of the configuration with respect to the adsorption energy does not change, being larger for O$_{fcc}$@Co/Ir and O$_{hcp}$@Co/Ir than for O@Co.

In Fig. S2 is shown the computed distance between the oxygen and the surface, for slabs with different number of layers. The values are very stable with fluctuations of 0.01 Å for the oxygen-surface distances.

### E. Computing the ground state magnetic moment

During the energy minimization we observed that the code sometimes arrives to an energy local minima, without finding the global minimum. In order to solve the problem, we perform several calculations with the total magnetization fixed and compute the total energy as a function of the value of the magnetization, as it is shown in Fig. S3. Choosing then the magnetic config-

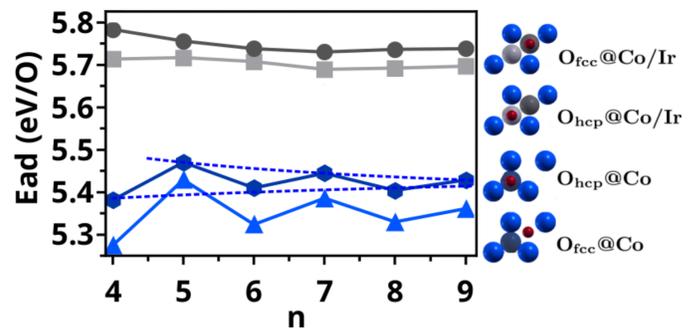

FIG. S1: Adsorption energy for an oxygen coverage of 25 ML on Co$_n$ (gray lines) and Co/Ir$_{n-1}$ (blue lines) slabs for the configurations shown on the right side of the plot. The A-B periodicity of the Co stacks is fitted with dashed lines in one of the cases in order to illustrated the convergence with the number of layers.

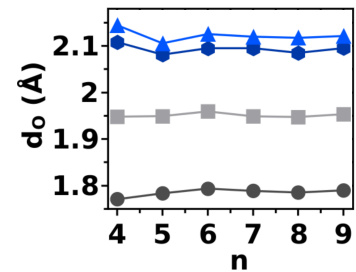

FIG. S2: Distance of O to the surface for the same systems in Fig. S1 (same color code), computed for on Co$_n$ and Co/Ir$_{n-1}$.

uration with the lower energy, we release the constraint on the magnetization and perform the structural relaxation. To reach the minimum, we often need to perform non-collinear magnetization calculations, even if the final results only show magnetization on the axis perpendicular to the surfaces.

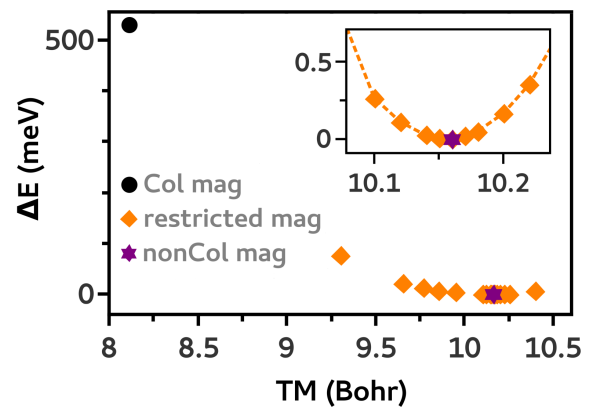

FIG. S3: Convergence of the ground state by sampling its total magnetization (TM).

| O (ML) | adsorption site | Co | Co/Ir |
|--------|-----------------|-----|-------|
| **0.5** | **hcp+fcc(octa)** | 2.95 [2] | |
| | **hcp+fcc(octa')** | 4.16 | 3.22 |

TABLE S4: Comparison of the adsorption energies for Co and Co@Ir obtained by considering O penetration. As it is explained in Section II, we did not find any $hcp + fcc(octa)$ solution, instead we find a configuration where the O atoms in the position $fcc$ are located under the second layer of the substrates.

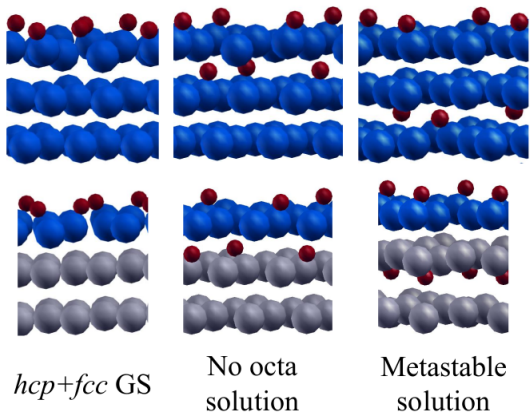

$hcp+fcc$ GS    No octa solution    Metastable solution

FIG. S4: Scheme illustrating the atomic structure corresponding to a) the ground state configuration with no subsurface O, b) the configuration found in Ref. [2], called $hcp + fcc(octa)$, and the only metastable solution with subsurface O (below the second layer) found in the present work.

## II. SUBSURFACE OXYGEN ADSORPTION

We have considered several starting configuration with oxygen atoms below the first Co layer. For both Co and Co/Ir surfaces, and for both O concentrations considered, upon geometry optimization the O migrates to the surface. We were not able to find any solution corresponding to the configuration $hcp+fcc(octa)$ as is define in Ref. [2]. This can be due to the different methods used to perform the structural relaxation. For example, in Ref. [2], in addition to oxygen, only the atomic positions of the atoms in the 3 top layers of a slab of 5 were relaxed, whereas we have relaxed the whole slab. Instead, we were able to compute a configuration with O occupying an $fcc$ site below the $2^{nd}$ layer. This corresponds to a Co layer for the case of the Co slab, but to an Ir layer for the case of Co/Ir and is the only configuration for which the adsorption energy is larger for Co than for Co/Ir. This configuration is considerably less stable then the surface sites, in fact, the interlayer positions are expected to occur only for O contents larger than 1 ML [2].

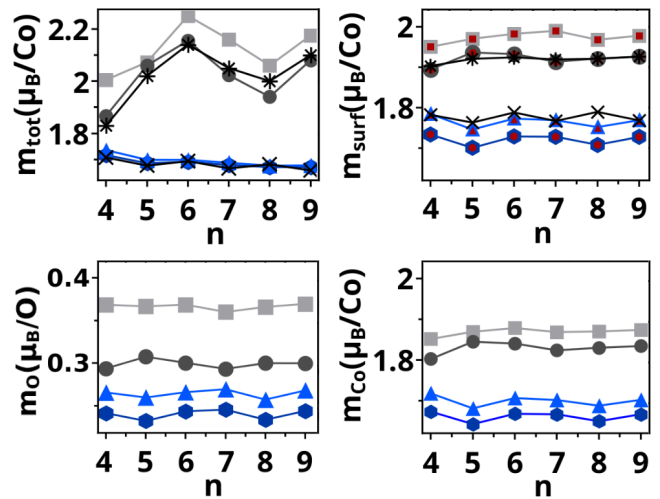

FIG. S5: Total slab magnetization (top-left) and oxygen (bottom-right), cobalt (bottom-right) and total (top-right) surface magnetization for Co@Ir and Co systems with 25 ML oxygen coverage computed for slabs with different number of layers. Colors are set according to Fig.S1, where we have filled the points with a smaller red zone to represent the sum in the total surface magnetization, $m_{Co} + m_O$. In the cases of the total slab and surface magnetization, the values for the pristine surface (black) are also shown. All the data is normalized to the number of Co atoms or to the number of O atoms for the oxygen magnetization.

## III. DEPENDENCE OF THE MAGNETIC PROPERTIES WITH THE SLAB THICKNESS

The dependence of the magnetic properties with the slab thickness for 0.25 oxygen ML is showed in Fig. S5. The total magnetization of cobalt systems with and without O converges to the bulk value, while the magnetization of Co/Ir systems varies between 1.8 and 2.3 $\mu_B$/Co. When O occupies the $fcc$ position, the magnetization is similar to one of the pristine surface, whereas when it occupies the $hcp$, the magnetization is higher by 0.1 $\mu_B$/Co.

## IV. AF METASOLUTIONS

CoO is know to have an antiferromagnetic order, as has been well described at the DFT level in Refs. [5–9]. This motivated the search for possible antiferromagnetic configurations for adsorbed O on the two Co surfaces considered previously.

We chose, among the configurations with higher O content, closer to the CoO stoichiometry, the ones with higher adsorption energies, labeled $fcc + fcc$ and $hcp + hcp$. In CoO, the antiferromagnetic pattern is formed by adjacent cobalt layers that are separated by an oxygen layer. In the present case, and since O is adsorbed on the surface, the only possibility for O to mediate an AF interaction is to consider configurations with in/plane AF coupling, as shown in Figure S6, although Co forms a tri-

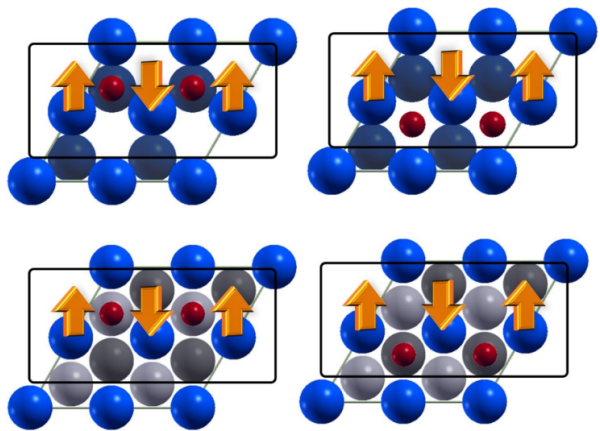

FIG. S6: Illustration of the different O adsorption sites (red dots) for a 0.5 ML concentration, with respect to the antiferromagnetic configuration imposed on the Co lattice (orange arrows). The upper panels represent O@Co and the bottom panels O@Co/Ir. Only half of the Co atoms present a non zero magnetic moment, and are arrange along a line in the $x$ direction.

|  | hcp+hcp | fcc+fcc |
|---|---|---|
| $E_{AF} - E_{FM}$ (eV/O) | 0.31 | 0.32 |
| $m_{tot}$ (abs) | 0.00 (5.07) | 0.00 (4.45) |
| $m_{Co}$ | ±1.70/0.00 | ±1.57/0.00 |
| $m_O$ | ±0.08/0.00 | ±0.07/0.00 |
| $m_{Ir}$ | ±0.10/0.00 | ±0.06/0.00 |

TABLE S5: Summary of the energy differences and magnetic moments corresponding to the AF configurations illustrated in Fig. S6. The reported energy correspond to the difference with respect to the ground state (ferromagnetic configuration). All the values of magnetization are given in $\mu_B$ per Co, O or Ir atom correspondingly.

angular lattice that will present frustration in the other Co atoms of the supercell. The Co rows along $x$ are coupled antiferromagnetically, following the distribution of the O atoms. The results are summarized in Table S5.

We find AF solutions only for Co/Ir. In case of Co no such configuration is found, since the strong ferromagnetic interaction in the slab imposed to the oxidized layer. In any case, the total energy of the AF configurations found for O@Co/Ir are 0.3 eV/O higher than the ferromagnetic solution.

In AF O@Co/Ir, the two rows of Co atoms closer to O show AF order. The other rows of Co atoms present zero magnetic moment as was expected. This is due to the fact that these atoms are surrounded by four Co atoms with different magnetic moment orientations, leading to magnetic frustration. It is possibly that the frustration be the reason making these configurations so unstable. The magnetization pattern of Co extends to the iridium layers: the Ir atoms close to magnetized Co planes are AF coupled, while the others have zero magnetic moment, resulting in a completely antiferromagnetic solution.





\end{document}